# Dual-Tuned $^{31}$P-$^{1}$H Dual-Row Loop/Dipole 32-element Transceiver Array for Human Brain Spectroscopy at 9.4T

G. A. Solomakha[1], R. Pohmann[1], F. Glang[1,2], S. Mueller[1], P. I. Valsala[1,3], T. Platt[4], S. Orzada[4], M. E. Ladd[4,5,6], A. Korzovski[4], K. Scheffler[1,7], N. I. Avdievich[1]

[1]High-Field MR Center, Max Planck Institute for Biological Cybernetics, Tuebingen, Germany

[2]Institute of Biomedical Imaging, Graz University of Technology, Graz, Austria

[3]WIN-Kolleg, Heidelberg Academy of Sciences and Humanities, Heidelberg, Germany

[4]Medical Physics in Radiology, German Cancer Research Center (DKFZ), Heidelberg, Germany

[5]Faculty of Medicine, University of Heidelberg, Heidelberg, Germany

[6]Faculty of Physics and Astronomy, University of Heidelberg, Heidelberg, Germany

[7]Department of Biomedical Magnetic Resonance, University of Tuebingen, Tübingen, Germany

Contact: g.solomakha@tuebingen.mpg.de

**Purpose**

The goal of this work is to develop and evaluate a single-layer tight-fit 32-element double-tuned loop/dipole transceiver (TxRx) array for human brain $^{31}$P MRS at 9.4T, achieving reasonable transmit and receive performance and full-brain coverage at both frequencies.

**Methods**

First, we developed numerical models of dual-row TxRx arrays for $^{31}$P (loop array) and $^{1}$H (coaxial-end folded-end dipole array) frequencies at 9.4T. Next, a multi-tissue voxel model was used to simulate Tx-performance of the arrays and define optimal CP-mode excitation. Following this, the proposed array performance was evaluated by MR measurements both on a phantom and a healthy volunteer. Finally, we compared the proposed array to a previously reported dual-tuned single-row loop-based TxRx array.

**Results**

The developed 32-element double-tuned array demonstrated full-brain (including the cerebellum and brain stem) imaging capabilities, reasonable SNR and transmit performance at both frequencies at 9.4T.

**Conclusion**

As a proof of concept, we developed a 32-element double-tuned UHF tight-fit TxRx human head array coil for $^{31}$P MRS with sufficient $^{1}$H performance using a combination of loop and dipole array elements. The proposed array design could also be adapted to higher fields, i.e., 10.5T, 11.7T, and 14T.

# 1 INTRODUCTION

MRS and MRSI using X-nuclei (other than $^1$H, for example – $^{19}$F, $^{31}$P, $^{23}$Na, $^{13}$C and $^2$H) are powerful methods to study brain metabolism (1). However, because of much lower concentrations, lower gyromagnetic ratios, and lower natural abundance in the human body (e.g. $^2$H and $^{13}$C) of non-zero spin X-nuclei, the achievable SNR is significantly worse than that of $^1$H. Therefore, X-nuclei MRI and MRS benefit from using ultra-high fields (UHF, $B_0 \geq 7T$) (1). Commonly, RF coils for X-nuclei and $^1$H are combined, resulting in so-called dual-tuned (DT) coils that are able to transmit and receive MR signals at both frequencies. Even in case of X-nuclei MRS(I) a $^1$H coil is typically required for providing spatial localization of X-nuclei data, anatomical reference, and static magnetic field shimming. In some cases, the $^1$H coil can also be used for proton decoupling (2,3) or/and polarization transfer (4,5) for $^{13}$C and $^{31}$P nuclei. In spite of the common RF designer's logic that the performance of a DT coil at $^1$H frequency is not that important and can be compromised on behalf of the X-nuclei performance, the $^1$H performance is critical for many applications. Having good performance at both frequencies in a DT coil allows to avoid moving a subject and repeating measurements using an additional single-tuned $^1$H coil, thus, substantially decreasing the total scanning time.

Due to the lack of vendor-provided transmit (Tx) body coils integrated into the bore of UHF MRI scanners, nested local Tx and receive (Rx) coils, which often consist of multi-element loop arrays, are used for both $^1$H and X-nuclei (6,7). Both Tx-only/Receive-only (ToRo) (8) and transceiver (TxRx) (6,9) array configurations can be used to design DT X-nuclei/$^1$H array coils. Using ToRo arrays at both frequencies, however, is problematic due to four interacting RF arrays, i.e., two Tx and two Rx arrays. To the best of our knowledge, a full-ToRo DT (i.e., both X-nuclei and $^1$H coils are ToRo) design has not been reported yet. Using TxRx arrays reduces the number of arrays to (10–12) (if only the $^1$H array is TxRx) or two (6,13) (if X-nuclei and $^1$H are both TxRx). Commonly, to simplify the two-array TxRx mechanical design, the $^1$H array is moved away a little further from the sample than the X-nucleus array, which decreases $^1$H SNR and overall image quality (6). On the other hand, placing both $^1$H and X-nuclei loop arrays in the same tight-fit layer, closer to the tissue, can preserve $^1$H performance without compromising SNR at the X-frequency as demonstrated in (14), where to enable placement of all $^1$H and X-nuclei loops in the same layer, we limited the total number of loops to 20 (10 at each frequency) by using single-row arrays. The developed DT array coil provided high Tx-efficiency at both frequencies without compromising SNR near the brain center at the $^{31}$P-frequency.

Using single-row tight-fit TxRx loop arrays at both frequencies, however, has several limitations: first, peripheral $^1$H SNR is not optimal and can be further improved by increasing the number of rows and the total number of loops, forming e.g. two rows of eight loops (2x8) (15). Additionally, for X-nuclei, increasing the loop number can also improve peripheral SNR at least for nuclei resonating at relatively high frequencies, e.g. $^{31}$P (162 MHz at 9.4T), at which the sample noise dominates (10,11). Second, at UHF, relatively large $^1$H-loops require a very large number of distributed capacitors to provide uniform voltage along the loops, e.g., 12 capacitors per single 100 mm by 100 mm loop, resulting in 96 for an 8-element TxRx array at 9.4T and almost 200 capacitors for a 2x8 array (15). Such a large number of capacitors significantly increases the array's complexity and decreases reliability. Finally, a single-row $^1$H loop array does not provide full-brain coverage at UHF (6,14,16). As demonstrated previously, the coverage can be further improved by using dual-row (2x8) loop arrays combined with 3D RF shimming (15). On the other hand, combining two 2x8 TxRx arrays for $^1$H and X-nuclei (32 loops in total) and placing both of them in the same layer greatly increases interaction between multiple loops and makes the overall design of the DT array very complex. To the best of our knowledge, a 32-loop TxRx DT loop array design has never been realized. In contrast to loops, dipoles can substantially simplify the human head array design (17–20). In DT array design, $^1$H dipoles can be combined with X-nuclei loops. Such a design, combining an $^1$H TxRx 8-dipole array with X-nuclei TxRx 8-loop array, was shown previously to provide relatively good proton imaging performance (21). In addition, dual-row (2x8) TxRx dipole arrays have been proposed for head (22) and for combined head and C-spine imaging (23) and showed good $^1$H imaging performance and RF shimming capability.

Based on the above-mentioned considerations, we developed and evaluated a DT 32-element single-layer loop/dipole TxRx array for $^{31}$P MRS at 9.4T by adapting a folded-end coaxial-end dipole design (23). The proposed array provides good $^{31}$P performance while offering superior $^1$H image homogeneity and full-brain coverage compared to the previously reported loop-based 20-element $^{31}$P/$^1$H array (14).

## 2 METHODS

### 2.1 Numerical Simulations

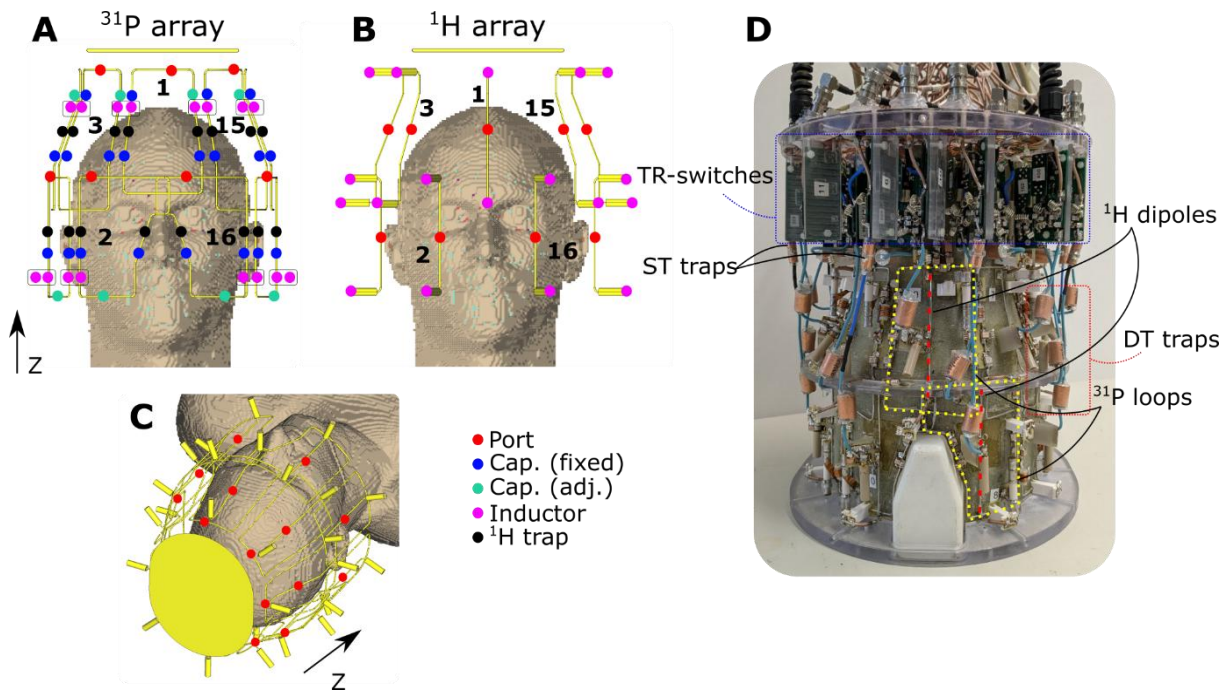

Figure 1. Top views of simulation models of the 162 MHz $^{31}$P-loop (A) and 400 MHz $^{1}$H-dipole (B) arrays in CST Studio 2021. (C) Isometric view of the 32-element DT array. (D) Photo of the constructed DT 32-element TxRx $^{31}$P/$^{1}$H array (FR-4 cover removed).

For the numerical simulations, we used the finite integration technique in the time domain, implemented in CST Studio 2021 (Dassault Systèmes, France) software. The proposed DT-array consisted of two rows (2x8) of coaxial-end folded-end $^{1}$H dipoles (400 MHz) (23) and two rows (2x8) of $^{31}$P loops (162 MHz) (i.e., 32 elements in total). In contrast to the previously reported 20-element TxRx loop array (14), which was used for comparison, the 32-element loop/dipole array was unshielded. Similar to previously reported dipole arrays developed in our group (20–22), only a local RF shield was placed 17 mm from the dipoles in the top row to enhance $B_1^+$ of $^{1}$H at the top of the head. Duke and Ella multi-tissue voxel models (24) cropped at the chest level were used for numerical simulations with tissue properties adjusted to the working frequencies. Because of the high complexity of the numerical models of the arrays, $^{31}$P and $^{1}$H arrays were simulated separately. Both loop and dipole array elements were constructed using 1.6 mm diameter copper wire with a 0.8 mm step local mesh applied to the coils' conductors. The numerical models of the proposed array loaded with the Duke voxel model are presented in 1A ($^{31}$P loop array) and 1B ($^{1}$H dipole array). The isometric view of the combined numerical models of both $^{1}$H and

$^{31}$P arrays is presented in Figure 1C. For tuning, matching, decoupling (for the loop array), and combining individual channel maps into CP-mode excitation, the CST Studio schematic module was used.

Array elements were placed on an elliptical (200 mm in width, 230 mm in height) FR-4 (glass fiber and epoxy) holder, established for TxRx arrays designs in our group (14,15,19,21,22,25). Longitudinal size of the $^{31}$P loops (along B$_0$ field direction, z-axis) was 110 mm in the bottom row and 125 mm in the top row. Adjacent $^{31}$P loops in the same row had 15-mm gaps between the conductors. They were decoupled using transformer decoupling (L=60 nH, K=-0.32), except for two elements in the bottom row near the nose, which were decoupled using overlapping. Adjacent elements in different rows were decoupled by overlapping (~30 mm). Along the conductors of the loop elements, four fixed capacitors were distributed in the top row and three in the bottom row. Each loop also had one tuning capacitor for fine-tuning the resonant frequency. Two $^1$H-frequency traps (band-stop filters) were placed in each $^{31}$P loop (Figure 1A) to minimize interaction between loops and dipoles at the $^1$H frequency. Series capacitive matching was used to match the input impedance of the loops to 50 Ohms. The $^1$H dipoles had a total length of 170 mm (130 mm along the z-axis and 2 × 20 mm coaxial ends). 50 Ohm cables with a central conductor diameter of 1.6 mm were used to construct coaxial ends. Coaxial ends were folded away from the sample to reduce sensitivity of the dipole resonance frequency to the sample size (23). At the coaxial end, a lumped 35-nH inductor was placed to provide a relatively uniform current distribution along the dipole conductor (25). An L-matching network consisting of a parallel capacitor and two series inductors was used to match the dipole elements to a 50 Ohm source. All arrays' models consisted of ~6·10$^7$ mesh cells, which resulted in a total simulation time using GPU acceleration (3xNVIDIA Tesla V100 32 GB) of 49 hours for the $^{31}$P array and 12 hours for $^1$H array both loaded with the Duke voxel model. Additionally, the $^{31}$P array was simulated using an elliptical phantom ($\varepsilon_{400}$=59, $\sigma_{400}$=0.64 S/m, $\varepsilon_{162}$=71, $\sigma_{162}$=0.65 S/m).

B$_1^+$ and local SAR over 10g of tissue (SAR$_{10g}$) for 1 W of stimulated RF power applied to the array inputs was calculated in the CP mode excitation (-45° shift b/w adjacent elements in one row, -22.5° b/w adjacent elements in different rows). The peak value of SAR over 10g of tissue SAR (pSAR$_{10g}$) was calculated using CST Legacy averaging methods. Tx-efficiency (<B$_1^+$>/√P) and spatial COV (measure of the homogeneity, std($B_1^+$)/<$B_1^+$>) were calculated over the brain region in numerical simulations at both frequencies. To improve $B_1^+$ homogeneity, the phase shift between the rows was optimized (15,22) at both frequencies. SAR efficiency was evaluated as <$B_1^+$>/$\sqrt{pSAR_{10g}}$. For comparison, a reference loop-only 20-element (10 TxRx $^1$H-frequency, 8 TxRx 2Rx $^{31}$P-frequency) DT array (14) was also simulated using the same numerical setup and the same metrics.

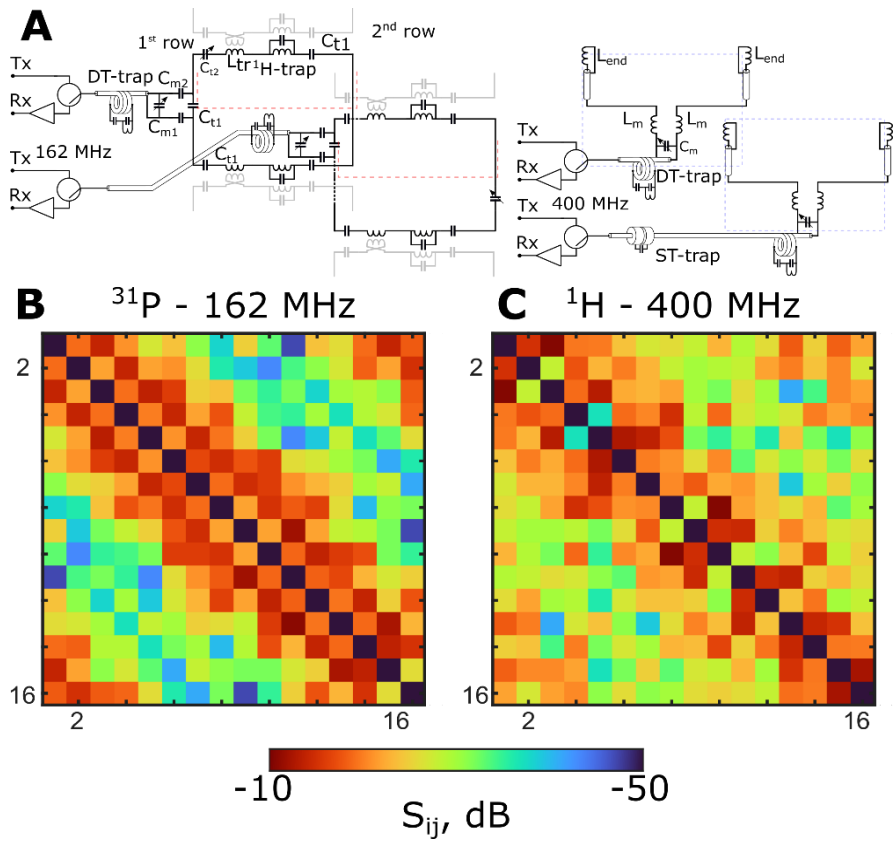

Figure 2. (A) Simplified electrical circuit of the 32-element TxRx $^{31}$P/$^1$H array. S-matrix measured on a healthy volunteer of the proposed array at 162 (B) and 400 (C) MHz.

**2.2 Array Construction**

After finishing numerical simulations, an array prototype was constructed and evaluated on the bench (Figure 1D). The simplified circuit of the constructed array is presented in Figure 2A. For both frequencies, high-power lumped-element 1:16 power splitters were designed with a phase shift corresponding to the optimal CP-mode excitation, defined by numerical simulations. Since the proposed array is TxRx, high-power home-built lumped-element PIN-diode TR-switches (16) in combination with OEM low-noise preamplifiers (WanTcom, MN, USA) were used at both frequencies. A combination of dual-tuned shielded cable traps (6) and floating-ground bazooka baluns (26) (only for dipole array elements) were used to eliminate common-mode propagation along the cable.

As for the numerical simulation, all array elements were constructed from 1.6-mm tin-coated copper wire (RS Group, England, UK). In the loop-array, high-power ceramic capacitors 100C (Kyocera-AVX, USA/Japan) were used ($C_{m1}$, $C_{t1}$, and $^1$H-trap). For matching, high-power variable capacitors ($C_{m1}$ and $C_{m3}$) 55H02 (Knowles-Johanson, USA) were used. For tuning the loop-array elements, high-power capacitors NMNT15 (Knowles-Johanson, USA) were used ($C_{t2}$). For dipole matching ($L_m$), dipole end-inductances ($L_{end}$), and $^1$H-traps, home-made inductors from tin-coated copper wire (1-mm diameter, Q≈120) were applied. A similar but enameled wire was used for decoupling transformers for in-row decoupling of loop elements ($L_{tr}$). Low-loss flexible K_02252_D-60 (Huber+Suhner, Switzerland) 50-Ohm cables were used for the coaxial ends. For shielded cable-trap construction and interconnection between array elements, inputs, and TR-switches, semi-flexible 50-Ohm non-magnetic RG-405 cable (McGill Microwave Systems, UK) was used.

Array elements were tuned and matched at both frequencies (162.2 MHz for $^{31}$P and 399.7 MHz for $^1$H) with an elliptical phantom for loading. The phantom was filled with water-sucrose-NaCl-solution ($\varepsilon_{400}$=58.3, $\sigma_{400}$=0.64 S/m, $\varepsilon_{162}$=71.4, $\sigma_{162}$=0.65 S/m). On-bench measurements were performed using a two-port VNA ENA 5063A (Keysight, CA, USA). The Q-factor for a single element of the $^{31}$P loop array was measured without $^1$H-traps, and with two $^1$H-traps using a decoupled double-probe (27) for loaded and unloaded cases. First, the array was tuned and matched on both frequencies using an elliptical homogeneous phantom. Next, by using the same elliptical phantom, we performed measurements of decoupling between $^{31}$P and $^1$H arrays at both frequencies by measuring the transmission coefficient ($S_{12}$) from the input of the $^{31}$P-splitter to the inputs of the $^1$H-array elements and by measuring $S_{12}$ from the input of the $^1$H splitter to the inputs of the $^{31}$P array elements. Prior to the in-vivo measurements, array element tuning and matching were adjusted at both frequencies on a healthy volunteer. Finally, full 16x16 S-matrices were measured at both frequencies (Figure 2B and C).

## 2.3 Experimental Evaluation of Array Performance

All measurements were performed at a 9.4T human whole body MR scanner (Siemens Magnetom Plus, Siemens Healthineers, Germany). $^{31}$P $B_1^+$ phantom measurement was done using the double-angle method (28) and CSI acquisition (TR=2400 ms, TE=0.35 ms, excitation FA=20°/40°, matrix size 28[HF] × 28[AP] × 32[LR], elliptical sampling, FOV 200 mm × 180 mm × 180 mm, Rx BW= 10 kHz, vector size 512). Because of the long TR of the $^{31}$P CSI sequence, the total measurement time was ~14 hours. The SNR in phantoms was determined based on another CSI acquisition (TR = 110 ms, TE = 0.35 ms, excitation FA = 40°, matrix size 48[HF]×48[AP]×48[LR], elliptical sampling, FOV 240 mm × 180 mm × 180 mm, Rx BW= 10 kHz, vector size 512) both for the proposed 32-element coil and 20-element reference loop arrays. $^{31}$P SNR was calculated for each voxel as the ratio of peak integral over the standard deviation of the noise part in the spectrum. After finishing the phantom experiment, an in-vivo measurement on a healthy volunteer was performed. The local ethics committee approved the in-vivo study, and informed consent was obtained from the subject before examination. $^{1}$H $B_1^+$ measurements were performed on both the 32- and 20-element arrays using AFI (TR = 20/100 ms, TE = 2 ms, excitation FA = 50°, matrix size 64[HF] ×64[AP] × 64[LR], FOV 224 mm × 224 mm × 224 mm, BW= 500 Hz/px) (29). Estimation of SNR in vivo was performed using a non-accelerated 3D gradient and RF-spoiled GRE (TR = 6 ms, TE = 3 ms, excitation FA = 5°, matrix size 220 [HF] × 216 [AP] × 176 [LR], 1 mm isotropic, BW= 450 Hz/px) dataset, combined with a noise-only dataset. The calculation of SNR in absolute units was performed offline using custom-written MATLAB scripts following the procedure described in (30). The AFI $B_1^+$ map was used for compensation of excitation inhomogeneity in the SNR maps. For in-vivo evaluation of $^{31}$P performance of the proposed array and comparison with the reference TxRx array, 3D k-space weighted CSI was used (TR = 70 ms, TE = 0.35 ms, excitation FA = 20°, matrix size 12 [HF] × 12 [AP] × 11 [LR], FOV 220 mm × 180 mm × 220 mm, Rx BW= 12 kHz, vector size 256). Processing included a WSVD (whitened singular value decomposition)-based coil combination (31) and a weighting correction to compensate for the integer-only k-space weighting. The missing spectral points at the beginning of each FID due to the time needed for phase encoding were extrapolated using Burg's method (32) before spatial zerofilling to 32 × 32 × 32 and Fourier-transform.

## 3 RESULTS

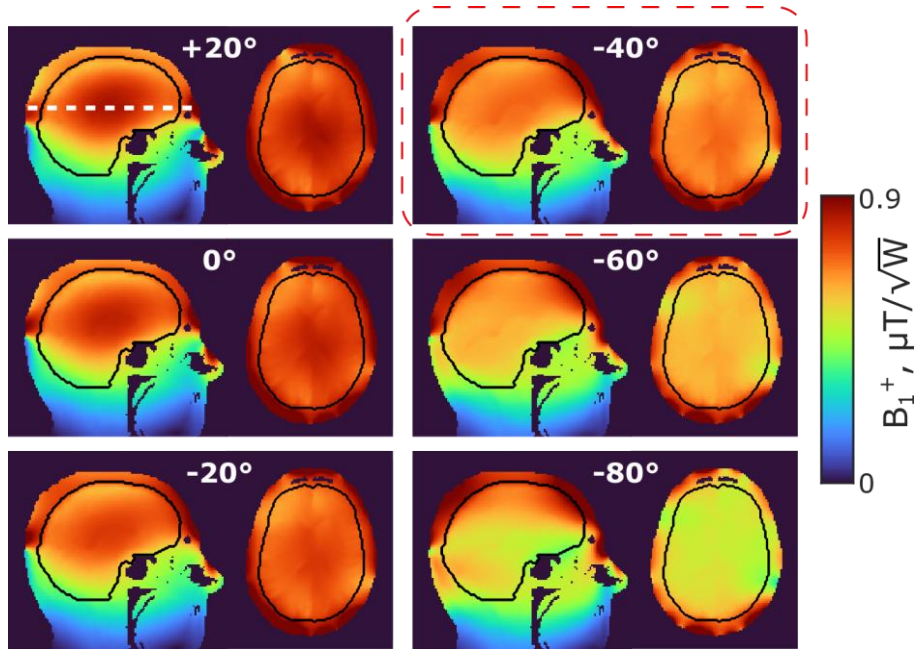

Figure 3. $B_1^+$ distributions simulated in the central sagittal slice and transversal slice through the maximum of $B_1^+$ of the Duke voxel model obtained for the dual-row 16-element $^{31}$P loop array with different phase shifts between the rows. Position of the transversal slice is marked with a white dash line. $B_1^+$ distribution with the optimal phase shift is marked with a red dashed rectangle. The region of the brain used to calculate <$B_1^+$>, COV, and SAR-efficiency is shown with a solid black line.

### 3.1 Numerical simulations

Figure 3 shows numerically simulated $B_1^+$ distributions in the sagittal (central) and transversal (through the $B_1^+$ maxima) planes of the Duke voxel model for the $^{31}$P double-row loop array with different phase shifts between the rows as indicated in the different subplots. The barplots in Figure 4 and Supp. Table 1 show a quantitative comparison of <$B_1^+$> (Tx-efficiency), COV, $pSAR_{10g}$, and SAR-efficiency for different phase shifts between the rows of the array for the Duke and Ella voxel models also in comparison to reference single-row 20-element array (8Tx 10Rx, referred as "Ref. loops"). For both voxel models, the dual-row array Tx efficiency drops with increasing phase shift between the rows. However, COV and $pSAR_{10g}$ show different trends compared to Tx-efficiency. The most homogenous excitation and lowest $pSAR_{10g}$ corresponds to -40° and -60° phase shifts. Similarly, the highest SAR-efficiency is achieved at similar phase shifts. As a compromise between homogeneity and Tx-efficiency, we chose the -40° phase shift between the rows. For Duke, the simulated Tx-efficiency at a -40° phase shift averaged across the brain was 0.633 µT/√W (6 % lower than the reference array), with a COV of 0.122 (20 % lower than the reference array). $pSAR_{10g}$ for -40° was 0.463 W/kg (25 % lower than reference array). This resulted in a SAR-efficiency of 0.929 µT/√W/kg (10 % higher than the reference array). In general, at the $^{31}$P frequency, the simulated Tx performance of the proposed array is very similar to that of the reference one.

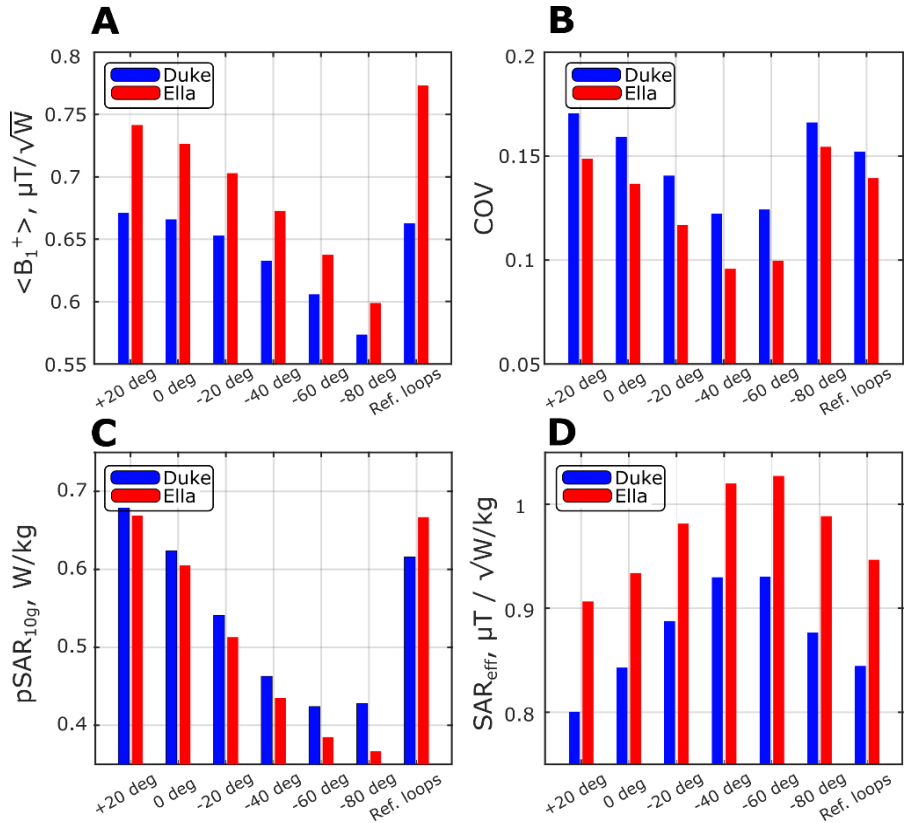

Figure 4. Bar plots showing $<B_1^+/\sqrt{P}>$ (A), COV (B), $pSAR_{10g}$ (C), and SAR-efficiency (D) obtained numerically for both the proposed 16-element $^{31}$P array (with different phase shifts between the rows) and the $^{31}$P part of the 20-element reference array (Ref. loops at figure) loaded by the Duke and Ella voxel models.

Figure 5 shows numerically calculated $B_1^+$ in the sagittal (central) and transverse (through the $B_1^+$ maxima) planes of the Duke voxel model for the $^1$H dipole array with different phase shifts between the rows. The coaxial-end dipole array provided full-brain excitation. Similar to the $^{31}$P array, adding a positive phase shift increased $B_1^+$ at the center, and adding a relatively small negative phase shift improved $B_1^+$ homogeneity. The bar plots in Figure 6 and Supp. Table 1 show a quantitative comparison of $<B_1^+>$, COV, $pSAR_{10g}$, and SAR-efficiency for different phase shifts between the rows of the dipole array for the Duke and Ella voxel models. The same metrics, as used for the $^1$H part of the reference single-row 20-element array (10TxRx, labeled "Ref. Loops" in the figures), are also presented in Figure 6 and Supp.

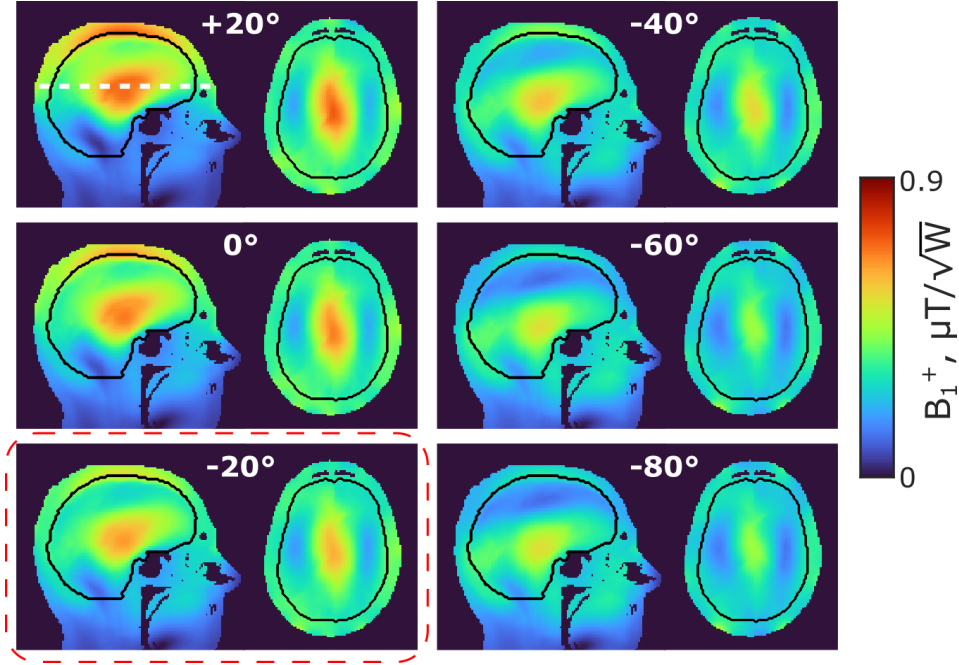

Figure 5. $B_1^+$ distributions simulated in the central sagittal slice and transversal slice through the maximum of $B_1^+$ of the Duke voxel model obtained for the dual-row $^1$H folded-end coaxial-end dipole array with different phase shifts between rows. Position of the transversal slice is marked with a white dash line. $B_1^+$ distribution with the optimal phase shift is marked with a red dashed rectangle. The region of the brain used to calculate <$B_1^+$>, COV, and SAR-efficiency is shown with a solid black line.

Table 1. The CP-mode with a -20° phase shift was chosen as a trade-off between Tx efficiency and homogeneity. Because of differences in the row count, array element count, and the individual element design (loops vs dipoles), the proposed and reference arrays showed significant differences in the simulated Tx performance. For the coaxial-end dipole array with a -20° phase shift, the Tx-efficiency over the brain was 0.294 µT/√W (28 % lower than for the reference array) with 0.301 COV (24 % lower than for the reference array). pSAR$_{10g}$ for -20° was 0.295 (67 % lower than for the reference array). This resulted in a SAR-efficiency of 0.541 µT/√W/kg (10 % higher than the reference array).

A comparison of numerically calculated $B_1^+$ distributions for the proposed array with optimal phase shift between rows and the reference arrays at both frequencies, for both arrays, and for the Duke and Ella voxel models is presented in Figure 7.

### 3.2 Experimental Evaluation of Array Performance

Full 16x16 S-matrices for the proposed array are presented in Figures 2B ($^{31}$P 162 MHz loop array) and 2C ($^1$H 400 MHz coaxial-end dipole array). S-matrix measurements were performed on the same subject, who later participated in in-vivo tests of the array. All elements were matched with S$_{ii}$ better than -13 dB. The worst measured coupling for the $^{31}$P array was -11.9 dB, measured between elements 11 and 1 in the top row. For the $^1$H array, the worst coupling was -10.6 dB, also measured in the top row, between elements 1 and 3. Insertion losses (above -12.04 dB of the ideal 16-way power splitter) of the 16-channel power splitters measured from the input interface cable to the input of the TR-switch were -0.6 dB for the $^{31}$P splitter and -1.15 dB for the $^1$H-splitter. Averaged measured S$_{12}$ from the input of the $^{31}$P splitter to the input of the $^1$H array elements were -32.3 dB (-28.6 dB worst, -35.5 dB best), averaged measured S$_{12}$ from the input of the $^1$H splitter to the input of the $^{31}$P array elements was -45.2 dB (-38.5 dB worst, -55.5 dB). Measured unloaded quality-factor (Q$_U$) were 298 and 211, and ratios of Q$_U$/Q$_L$ (unloaded to loaded quality factor) was 3.87 and 3.35 for the no-trap and two-trap loops, respectively. Based on the analysis proposed in (33), the insertion of two traps decreases SNR by 5.7 % compared with the loop without traps.

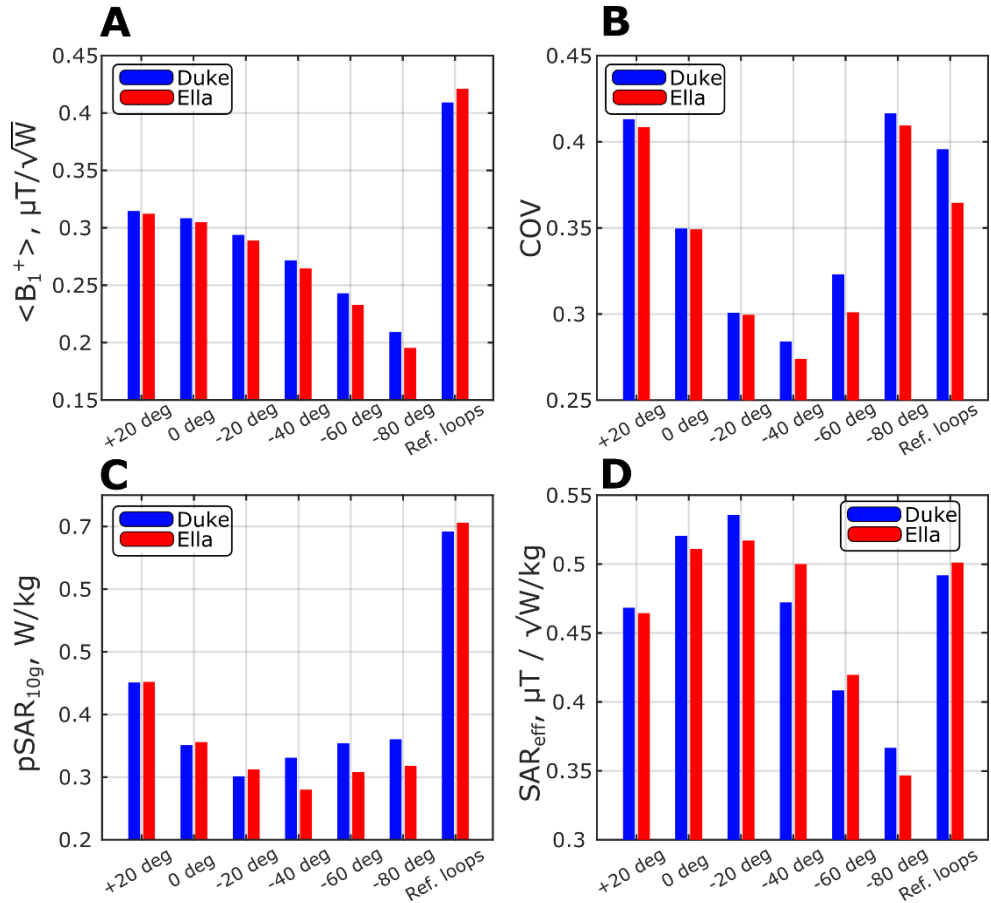

Figure 6. Bar plots showing <$B_1^+/\sqrt{P}$> (A), COV (B), pSAR$_{10g}$ (B), and SAR-efficiency (C) obtained numerically for both the proposed 16-element $^1$H array (with different phase shifts between the rows) and the $^1$H part of the 20-element reference array (Ref. loops at figure) loaded by the Duke and Ella voxel models.

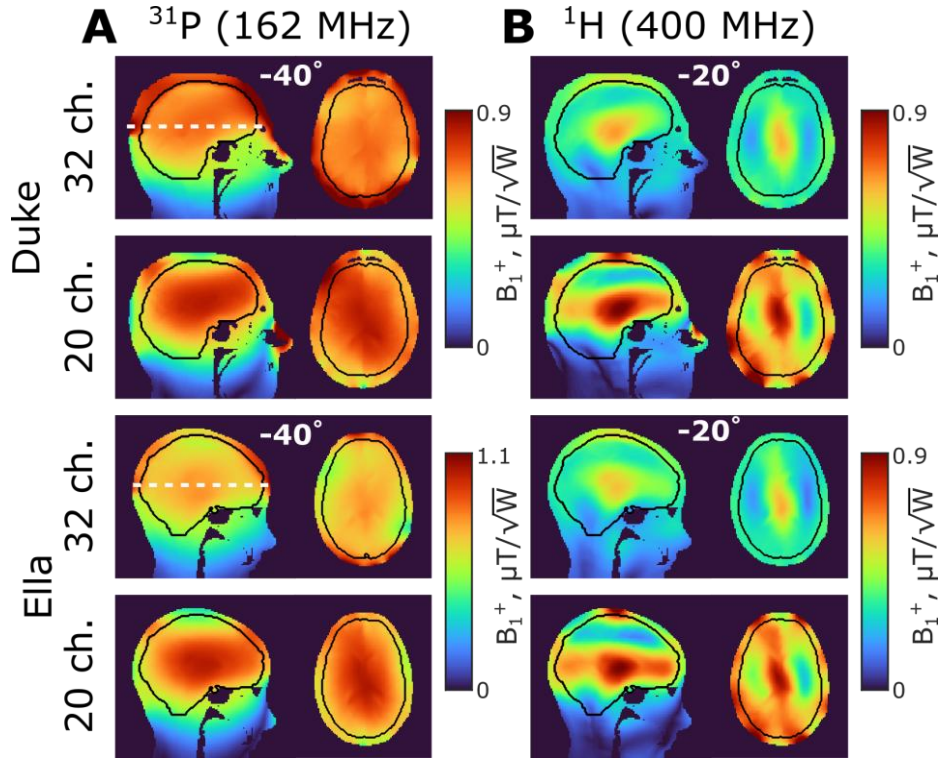

Figure 7. $B_1^+$ distributions simulated in the central sagittal slice and transversal slice through the maximum of $B_1^+$ of the Duke and Ella voxel models obtained at $^{31}P$ (A) and $^{1}H$ (B) frequencies for the proposed 32-element (the optimal phase shift between the rows) and reference 20-element arrays. Position of the transversal slice is marked with a white dash line. The region of the brain used to calculate <$B_1^+$>, COV, and SAR-efficiency is shown with a solid black line.

Measured (double-angle method) and numerically simulated $B_1^+$ distributions in the elliptical phantom at the $^{31}P$ frequency are presented in Figure 8A. This figure demonstrates, we can see good agreement in the magnitude of $B_1^+$ in the phantom center between measurement and simulation. Measured $^{31}P$ CSI SNR for 32- and 20-element arrays are presented in Figure 8B. In general, the 32-element array showed $^{31}P$ receive performance similar to that of the previously reported 20-element array (reference one (14)), with a slight improvement in peripheral SNR. The 20-element array showed better SNR at the top (head direction) of the phantom because of the additional two Rx-only vertical loops (14).

In-vivo $^{1}H$ $B_1^+$ field distributions are presented in Figure 9A. Similarly to numerical simulation, the reference loop array produces higher $B_1^+$, but with significantly lower spatial homogeneity and worse coverage over the brain region. Measured in-vivo $^{1}H$ SNR maps of both arrays are presented in Figure 9B. Measured in-vivo $<B_1^+>$ over a 150-mm slab covering the whole brain (shown with a white rectangle in Figure 9A) and mean SNR in two ROIs (ROIs in Figure 9B) are presented in Supp. Table 2. Mean measured $<B_1^+>$ in the slab of the volunteer's brain was 0.264 µT/√W for the 32-element array (23 % lower than for the 20-element array). As seen in Figure 9B, the SNR of the dipole array is lower than that of the reference array at the periphery of the brain (27 % decrease of mean SNR for the 32-element array compared to the 20-element array in ROI$_1$), while near the center the SNR (ROI$_2$) is very similar (difference of 2.3 %). Anatomical GRE-images of a healthy volunteer are presented in Figure 10A. In-vivo $^{31}$P SNR maps for both arrays are presented in Figure 10B. $^{31}$P spectra from two different locations in the human brain, acquired using the 32- and 20-element arrays, are presented in Figure 10C. Similar to the phantom measurements, the 32-element array provides slightly better SNR at the periphery, while the 20-element array provides better SNR at the top of the head due to the additional Rx-only elements there.

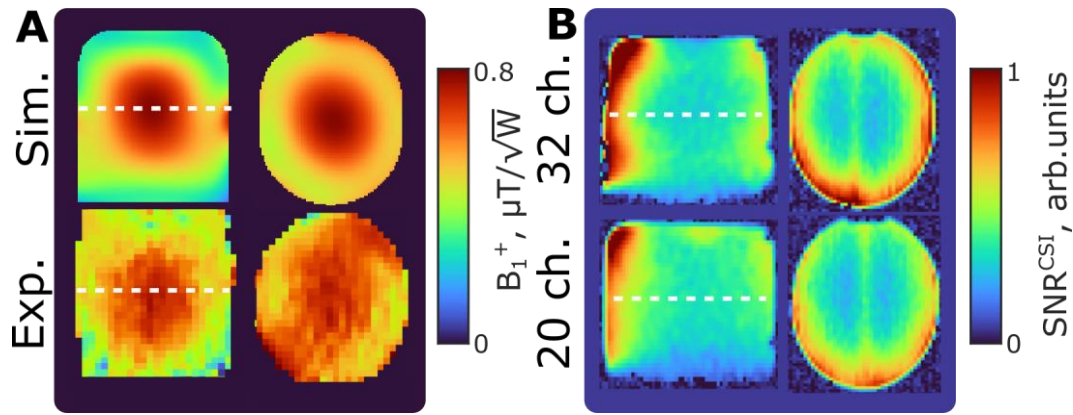

Figure 8. (A) Simulated and measured $B_1^+$ in the homogenous elliptical phantom in the central sagittal and transversal slice through the maximum of $B_1^+$. (B) Measured CSI-SNR in the homogenous elliptical phantom in the central sagittal and transversal slices. Position of the transversal slice is marked with a white dash line.

## 4 DISCUSSION

We developed, constructed, and evaluated, both numerically and experimentally, a novel $^{31}$P/$^{1}$H (162/400 MHz, 9.4T) 32-element combined loop/dipole TxRx array design for brain imaging and spectroscopy. To provide good loading, and therefore good performance at both frequencies, both the X-nuclei and $^1$H arrays were placed in the single layer on the surface of a tight-fit housing, similarly to previously reported DT TxRx arrays (14,21). Placing all 32 elements in a single layer was made possible due to the simplicity of the dipole design which, unlike loops, does not require distributed capacitors. Furthermore, following our previous development (23), we utilized folded-end coaxial-end dipoles, which allowed minimizing the frequency shift under different loading conditions as, e.g., due to variations in head size (34). The latter is a characteristic feature of the common dipoles, which are very difficult to use in tight-fit TxRx or Rx arrays (34).

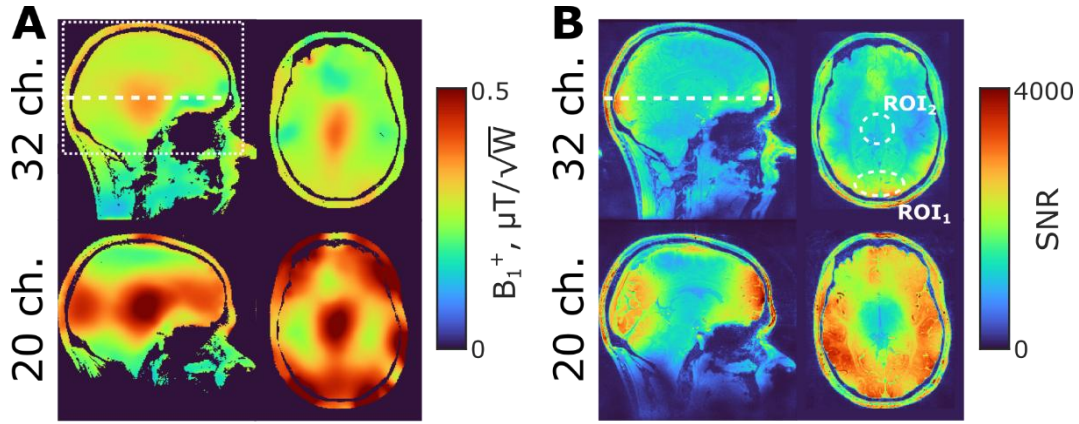

Figure 9. (A) In-vivo measured $B_1^+$ in the central sagittal and transversal slice through the maximum of $B_1^+$. Region used for $<B_1^+>$ calculation shown with a white dashed rectangle. (B) Measured in-vivo SNR in central sagittal and transversal slices through the maximum of $B_1^+$. ROIs used for mean SNR calculation shown with dashed ellipses. Position of the transversal slice is marked with a white dash line.

In spite of such a densely populated design, we were able to maintain sufficient decoupling between all array elements resonating at the same frequency. Relatively short coaxial-end folded-end dipoles are intrinsically decoupled. While the X-nuclei loop array required decoupling between adjacent loops (transformer decoupling within the rows and overlapping between the rows), the $^1$H dipole array did not require any additional decoupling. Loops and dipoles resonating at different frequencies were also well decoupled. No X-nuclei cable traps were required in the dipoles, since dipoles and loops were geometrically decoupled. According to our Q-factor measurements, this only slightly reduced the $^{31}$P SNR (5.7 % SNR reduction).

As a result, the developed 32-element dual-row DT array substantially improved the transmit homogeneity and coverage at the $^1$H frequency as compared to the reference 20-element single-row array. Simulated $^1$H $B_1^+$ COV obtained for the proposed dipole array with -20° phase shift between the rows was 24 % lower than that of the reference loop array. As demonstrated previously (15,22), multi-row arrays combined with 3D RF shimming are necessary for providing good homogeneity and transmit coverage across the whole brain and sufficient homogeneity. As seen in Figures 5 and 6, even a relatively simple RF shimming (phase shift between the rows) results in substantial improvement of homogeneity and coverage. Simulations agree well with the experimental data. As seen in Figure 9, the proposed array provides full-brain coverage in transmit, including the cerebrum, cerebellum, and brain stem while the reference array has a very limited coverage. However, measured at 400 MHz the dipoles' Tx-efficiency was 23 % lower than that of the loop-based reference array (Supp. Table 1). At the same time, the proposed array demonstrated a 67% decrease in $pSAR_{10g}$, resulting in a 10 % increase in the SAR-efficiency. Similar to the numerical simulation, the measured $<B_1^+>$ in the volunteer's brain was lower with the 32-element array for the chosen ROIs compared to the 20-element array. This, in general, confirms previous results (35,20,36) showing superior peripheral Rx-performance of the loop array over dipoles at 7 and 9.4T. Comparison of the central SNR also agrees with predictions of Ultimate Intrinsic SNR (UISNR) theory (37) showing increasing contribution of the dipoles with increasing $B_0$ field strength with 9.4T being approximately the point of equality. Full-brain coverage and reasonable SNR of the DT array at the $^1$H frequency is crucial for providing a high-quality anatomical reference for the co-registration of X-nuclei MRSI data and for achieving good $B_0$-shimming across the whole ROI. Also, a good $^1$H performance allows for obtaining good anatomical data without switching from X-nuclei to a dedicated imaging coil.

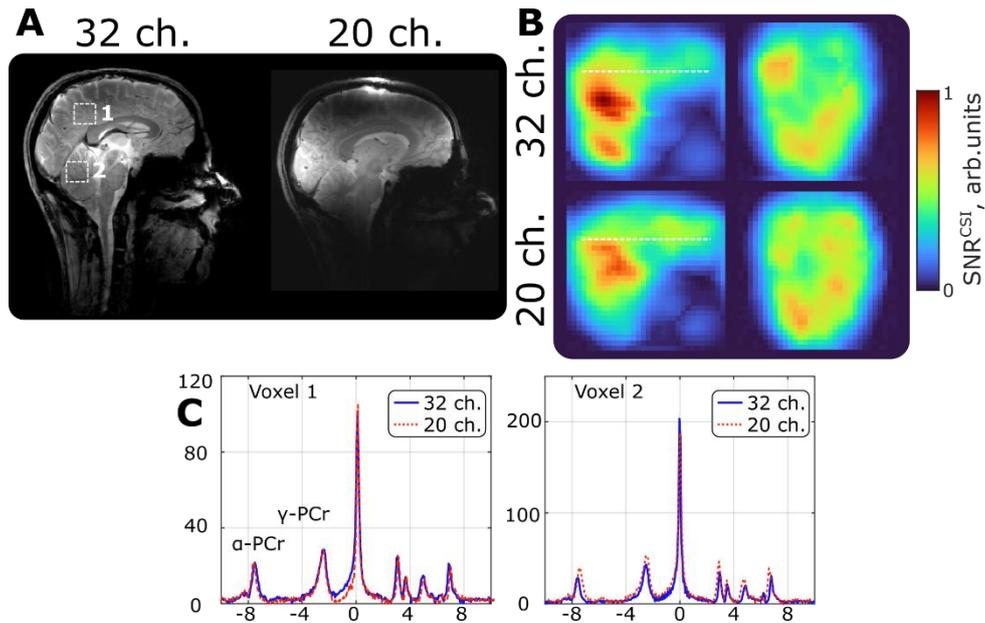

Figure 10. (A) In-vivo anatomical GRE images obtained using the proposed and reference arrays. (B) $^{31}$P in-vivo CSI-SNR for the proposed 32- and reference 20-element arrays in a central sagittal (left column) and transversal slice (right column, position indicated by white dashed line). (C) In-vivo $^{31}$P spectra.

Regarding the performance at the $^{31}$P frequency (Figure 8B) the proposed array provides a slight improvement in peripheral SNR while maintaining similar performance in the phantom's center. Further measurements confirmed this for the in-vivo case. Also, the proposed array shows an improvement of SNR in the cerebellum. Since the proposed array has two rows of elements, the distribution of the Tx field at 162 MHz can be controlled by adjusting the phase shift between the rows of the TxRx array. It is noteworthy that the newly developed and the reference array have some difference in the design in addition to the number of rows: the reference array has two additional Rx-only loops placed at the top of the head. As seen in both phantom and in-vivo measurements (Figures 8B and 10B), the reference array showed higher SNR at the top of the imaging object. The presence of these elements most likely also influences the central SNR. Due to the limited number of Rx channels at our scanner (max. 32), placement of such elements, unfortunately was not possible in the presented array. This, however, may become feasible for UHF systems with more receive channels.

To avoid degradation of the SNR due to the intrinsic coil noise, it is important to maintain sample noise domination regime (33), which is defined by the ratio of $Q_U/Q_L$. For the $^{31}$P loops in the proposed array, measured $Q_U/Q_L$ was 3.35 corresponding to ~16% decrease of the SNR in comparison to the ideal case a lossless RF coil (33). Adding $^1$H traps further decreased SNR by ~6%. Since sample losses decrease with decreasing frequency and loop size, it is harder to maintain the sample noise domination regime. This makes the use of dual-row loop-based TxRx arrays problematic at lower frequencies. Hence, we expect that the proposed array architecture (16TxRx X-nuclei loops combined with 16TxRx $^1$H dipoles) could be promising for DT-array design for brain imaging in static fields even higher than 9.4T. For example, at 11.7T and 14T, the $^{31}$P frequency reaches 201 and 241 MHz, and the $^{23}$Na and $^{13}$C frequencies are ~130 and ~160 MHz. The higher X-nuclei frequency will result in stronger $B_1^+$ inhomogeneity, especially for $^{31}$P, where a double-row array design will enable optimizing of $B_1^+$ distribution and minimizing pSAR.

## 5 CONCLUSION

We developed a 9.4T DT-array with 32 TxRx elements (16 [1]H coaxial-end folded-end dipoles and 16 [31]P loops), all placed in a single layer. Our results show reasonable transmit and receive performance at both frequencies, with significant improvements in [1]H coverage and SAR efficiency, and higher [31]P peripheral SNR, as compared with a previously developed 20-element single-layer loop-only array. The proposed array design could be easily adapted to $B_0$ higher than 9.4T, i.e., 10.5, 11.7, and 14T, where dipoles would show better $B_1^+$ efficiency and higher SNR compared to loops.

## 6 ACKNOWLEDGEMENTS

Funding by the European Research Council (ERC Advanced Grant No 834940, SpreadMRI) and DFG HF-NeuroBOOST (DFG/ANR joint project; project number 530130666) is gratefully acknowledged. Dr. Loreen Ruhm is gratefully acknowledged for help in setting up the [31]P SNR phantom measurements. Tim Haigis is gratefully acknowledged for help in setting up in-vivo $B_1^+$ and SNR measurements.

Supp. Table 1. Numerically calculated Tx efficiency (mean $B_1^+$ over the brain, normalized to the square root of the stimulated power), $pSAR_{10g}$, SAR-efficiency (mean $B_1^+$ over the brain normalized to the square root of $pSAR_{10g}$), and COV (standard deviation of $B_1^+$ normalized to mean $B_1^+$ over the brain) for the Duke voxel model for the 32- and 20-element arrays.

| Array | Row phase shift, ° | $<B_1^+>$, µT/√W | COV | $pSAR_{10g}$, W/kg | $SAR_{eff}$, µT/√W/kg |
|---|---|---|---|---|---|
| 162 MHz – $^{31}P$ | | | | | |
| 32-element array | 20 | 0.671 | 0.171 | 0.703 | 0.801 |
| 32-element array | 0 | 0.667 | 0.159 | 0.624 | 0.843 |
| 32-element array | -20 | 0.653 | 0.141 | 0.541 | 0.889 |
| **32-element array** | **-40** | **0.633** | **0.122** | **0.463** | **0.929** |
| 32-element array | -60 | 0.601 | 0.124 | 0.424 | 0.931 |
| 32-element array | -80 | 0.574 | 0.166 | 0.428 | 0.877 |
| 20-element array | - | 0.663 | 0.152 | 0.616 | 0.845 |
| 400 MHz – $^1H$ | | | | | |
| 32-element array | 20 | 0.315 | 0.413 | 0.451 | 0.468 |
| 32-element array | 0 | 0.381 | 0.349 | 0.351 | 0.520 |
| **32-element array** | **-20** | **0.294** | **0.301** | **0.295** | **0.541** |
| 32-element array | -40 | 0.271 | 0.284 | 0.331 | 0.472 |
| 32-element array | -60 | 0.243 | 0.323 | 0.354 | 0.408 |
| 32-element array | -80 | 0.211 | 0.420 | 0.357 | 0.352 |
| 20-element array | - | 0.409 | 0.395 | 0.692 | 0.491 |

Supp. Table 2. Measured Tx-efficiency (mean $B_1^+$ over the brain, normalized to the square root of the stimulated power), and SNR at the $^1H$ frequency in the healthy volunteer for the 32- and 20-element arrays.

| Array | $<B_1^+>$, µT/√W | Mean $SNR_{ROI1}$ | Mean $SNR_{ROI2}$ |
|---|---|---|---|
| 32-element array | 0.264 | 2038 | 1197 |
| 20-element array | 0.343 | 2803 | 1225 |